\documentclass[12pt]{article}
\pdfoutput=1
\usepackage{comment}
\usepackage{soul}
\usepackage{geometry}
\usepackage{amsmath, amssymb, bbm, amsthm, mathabx, bigints}

\usepackage{float}
\usepackage{setspace}
\usepackage[font=small]{caption}
\usepackage[colorlinks,citecolor=DarkBlue,linkcolor=DarkBlue,bookmarks=false,hypertexnames=true, urlcolor=blue]{hyperref} 
\usepackage[svgnames]{xcolor} % this package has dark bluefor hyperlinks
\usepackage{stfloats} % placing tables
\usepackage{enumitem}

\usepackage[affil-it]{authblk} % edit font size of author on cover page 
\usepackage{bbm}

\usepackage{fancyhdr}
\usepackage[font=footnotesize,labelfont=sc]{caption}
\usepackage{accents}
\renewcommand{\thesection}{\Roman{section}} 
\usepackage{titlesec}
\usepackage{lipsum}
\usepackage{lmodern}

\renewcommand{\thesection}{\arabic{section}}
\renewcommand{\thesubsection}{\thesection.\arabic{subsection}}
\renewcommand{\thesubsubsection}{\thesubsection.\arabic{subsubsection}}

\titleformat{\section}
  {\centering\scshape\normalsize}
  {\thesection. }{0.1em}{}

\titleformat{\subsection}{\centering\normalsize\itshape}{\thesubsection\ }{0.1em}{}

\titleformat{\subsubsection}[runin]{\normalsize\itshape}{\thesubsubsection}{1em}{}

\usepackage{graphicx}
\usepackage{grffile}
\usepackage{tablefootnote}
\usepackage{dcolumn}
\usepackage{tabularx, afterpage}
\usepackage{pdflscape}
\usepackage{pdfpages}

\usepackage{footnote}
\makesavenoteenv{tabular}
\usepackage[multiple]{footmisc}

\usepackage[toc,page, title]{appendix}

\usepackage{csquotes}

\usepackage{natbib}

\usepackage{dsfont,booktabs,makecell,cleveref,tikz}

\renewenvironment{abstract}
 {\small
  \begin{center}
  \bfseries \abstractname\vspace{-.5em}\vspace{0pt}
  \end{center}
  \list{}{%
    \setlength{\leftmargin}{21mm}% <---------- CHANGE HERE
    \setlength{\rightmargin}{\leftmargin}%
  }%
  \item\relax}
 {\endlist}

\newtheorem{proposition}{Proposition} 
\newtheorem{assumption}{Assumption}  
\newtheorem{lemma}{Lemma}

\usepackage{chngcntr}
\usepackage{apptools}
\AtAppendix{\counterwithin{lemma}{section}} % restart counter for appendix
\AtAppendix{\counterwithin{assumption}{section}}

\begin{document}

\makeatletter
\patchcmd{\@maketitle}{\LARGE \@title}{\fontsize{12}{12}\selectfont\@title}{}{}
% custom fontsize
\newcommand\customsize{\@setfontsize\customsize{8.5}{11}}
\makeatother

\renewcommand\Authfont{\fontsize{12}{12}\selectfont} % size of author font
%\renewcommand{\abstractname}{}    % clear the abstract title

% footnote without marker
\newcommand\blfootnote[1]{%
  \begingroup
  \renewcommand\thefootnote{}\footnote{#1}%
  \addtocounter{footnote}{-1}%
  \endgroup
}

\font\myfont=cmr12 at 16pt
%\title{{\LARGE Optimal Screening in Experiments with Partial Compliance\thanks{
%We thank \hl{[ADD]} and seminar participants at the Australian National University, Deakin University, Oxford %University, University of Sydney, University College London \hl{[ADD]} for helpful comments.
%}}}
%\title{{\LARGE Optimal Screening in Experiments with Partial Compliance}}

\title{{\LARGE Optimal Screening in Experiments with Partial Compliance\thanks{
We thank Peter Hull and seminar participants at the 2025 STEP UP Conference at UNSW for helpful comments.
}}}

\author[1]{\Large Christopher Carter\thanks{Department of Economics, University of Technology Sydney. Email: \href{christopher.carter@uts.edu.au}{christopher.carter@uts.edu.au}} \; \;
Adeline Delavande\thanks{Department of Economics, University of Technology Sydney. Email: \href{adeline.delavande@uts.edu.au}{adeline.delavande@uts.edu.au}} \; \;
Mario Fiorini\thanks{Department of Economics, University of Technology Sydney. Email: \href{mario.fiorini@uts.edu.aue}{mario.fiorini@uts.edu.au}} \; \;
Peter Siminski\thanks{Department of Economics, University of Technology Sydney. Email: \href{peter.siminski@uts.edu.au}{peter.siminski@uts.edu.au}} \; \;
Patrick Vu\thanks{Department of Economics, University of New South Wales. Email: \href{patrick_vu@brown.edu}{patrick\_vu@brown.edu}}}
\date{\today}

% --- make symbols for title/author footnotes ---
\maketitle
\renewcommand{\thefootnote}{\arabic{footnote}}
\setcounter{footnote}{0}

\begin{abstract}
This note studies optimal experimental design under partial compliance when experimenters can screen participants prior to randomization. Theoretical results show that retaining all compliers and screening out all non-compliers achieves three complementary aims: (i) the Local Average Treatment Effect is the same as the standard 2SLS estimator with no screening; (ii) median bias is minimized; and (iii) statistical power is maximized. In practice, complier status is unobserved. We therefore discuss feasible screening strategies and propose a simple test for screening efficacy. Future work will conduct an experiment to demonstrate the feasibility and advantages of the optimal screening design. 
\end{abstract}

\thispagestyle{empty}

\newpage
\section{Introduction}
In experiments with partial compliance, low take-up rates can make instrument variables (`IV') estimates of the average treatment effect on the treated both biased and imprecise. In response, recent econometrics research has proposed alternative methods aimed at improving precision \citep{Spiess2021, Hazard2025}. Such efforts are based on the insight that placing greater weight on complier populations can enhance efficiency. For example, \citet{Spiess2021} propose an alternative estimator that weighs each observation according to its estimated compliance; and \citet{Hazard2025} exploits covariate information to restrict estimation to a subpopulation with non-zero compliance. 

This note studies a related, but distinct, problem. It considers optimal experimental design under partial compliance when experimenters can screen participants prior to randomization. This differs from previous studies, which propose alternative estimators that can be applied to existing IV studies \citep{Spiess2021, Hazard2025}. By contrast, this note addresses experimental design.

Our main aim is to establish the conditions under which a screened 2SLS estimator dominates the standard 2SLS estimator in terms of both bias and power. We study this question in the standard setup where both the instrument and treatment are binary. To begin, we consider an idealized setting where the experimenter observes participant types (i.e. compliers, always-takers, and never-takers). In later sections, we discuss how types can be elicited in practice. 

A first-order concern is that screening can change the target parameter. This can render comparisons between screened and unscreened estimators ambiguous. Thus, to begin, we establish a sufficient condition which guarantees that the screened 2SLS estimator identifies the same Local Average Treatment Effect (`LATE') as the unscreened 2SLS estimator. Specifically, the sufficient condition states that any screening mechanism which retains all compliers identifies the same LATE. The intuition is simple: because the LATE only identifies the treatment effect for compliers, any screening mechanism retaining all compliers must preserve the target parameter.

Under this sufficient condition, we then show that the optimal screening mechanism is to retain all compliers and screen out all non-compliers (i.e. always-takers and never-takers). This screening mechanism is `optimal' in the sense that it minimizes bias and maximizes statistical power among the set of screening mechanisms that retain all compliers. Given that this set also includes no screening, the optimal screened 2SLS estimator therefore dominates the standard, unscreened 2SLS estimator in terms of both bias and power. Intuitively, the 2SLS estimator only identifies the average treatment effect among compliers, so that non-compliers only contribute noise to the estimator. In effect, removing non-compliers therefore `strengthens' the instrument; and this more than offsets decreases in precision due to a smaller sample size. It is noteworthy that the optimal screened estimator does not exhibit a bias-variance trade-off, which can be can be present in alternative approaches \citep{Spiess2021, Hazard2025}, as both bias and efficiency improve. 

%Moreover, screening is likely to come with lower costs for the experimenter, for the simple reason that data need only to be collected for the subsample that passes the screening stage. Thus, if the optimal estimator is feasible, then it comes with potentially substantial gains and lower costs. 

The theoretical results rest on the assumption that the experimenter observes participants' types, which is not true in practice. Thus, to practically implement the screening estimator, the experimenter must be able to accurately elicit types. To do this, we suggest that experimenters simply ask participants two questions, with the proviso that their answers will have no bearing on whether they will be offered treatment: (i) \textit{``Would you take treatment $(D_i=1)$ if offered $(Z_i=1)$?''} and (ii) \textit{``Would you take treatment $(D_i=1)$ if not offered $(Z_i=0)$''}?\footnote{In an RCT with one-sided non-compliance, it is only necessary to ask participants question (i) because treatment is not available to those not offered.} 

If reported compliance matches actual compliance, then types are accurately identified and the optimal screening estimator is feasible. To test whether this is the case, we propose an approach based on results in \citet{Hull2024}, which allow one to estimate the probability that stated complier status equals one conditional on actually being a complier.\footnote{More formally, let $D_i$ be binary treatment, $CM_i$ be a binary variable for \textit{true} complier type, and $\widetilde{CM}_i$ be a binary variable that equals one for reported compliers. Results in \citet{Hull2024} show that an IV regression (with the full sample) of $\widetilde{CM}_i \times D_i$ on $D_i$ estimates $\mathbbm{P}[ \widetilde{CM}_i=1 | CM_i=1]$.} 

This provides a test of the identification assumption that all compliers are retained under screening, and has implications for which estimator should be reported. For example, if the null hypothesis that $\mathbbm{P}[ \widetilde{CM}_i=1 | CM_i=1]=1$ is rejected, then researchers should use the standard, unscreened estimator because the screened estimator identifies a different LATE. Alternatively, failing to reject the null provides support for the use of the screened estimator, which will identify the same LATE and have better statistical properties as compared with the unscreened estimator in terms of both power and bias.

Finally, it is important to add that we recommend that experimenters actually implement `pseudo-screening', where they ask the screening questions, but retain the full sample for the experiment. This allows them to calculate both the screened and unscreened estimator, estimate the power gain from the screened design, and also perform stronger tests of the identification assumption, as discussed in more detail in the main text.

\begin{comment}
Whether or not this is the case is \textit{partially} testable, by observing if participants' \textit{reported type} match \textit{actual type} based on their choices in the experiment. To illustrate, consider the simple case of an RCT with one-sided noncompliance where control units cannot access treatment (i.e. $Z=0 \Rightarrow D=0)$. In this setting, there exist only compliers and never-takers. Now consider a participant whose \textit{reported type} in the screening stage is a complier. If they are offered treatment in the experiment ($Z=1$), then their observed treatment choice reveals their \textit{true type}. Hence, whether their \textit{reported type} matches their \textit{true type} identifies the truthfulness of their initial response. On the other hand, if the participant is not offered the treatment ($Z=0$), then the experimenter never observes their true type and cannot know whether their initial report was truthful.

Partial testing of identification assumptions is not unique to the screening problem. For example, balance tests in RCTS or RDDs are able to verify no statistical differences in observed characteristics, but are silent on balance across unobserved characteristics. Similarly, in difference-in-differences studies, parallel pretrends provide support for the identification assumption but does not -- and cannot -- verify parallel trends in the post-treatment period in the absence of treatment.
\end{comment}

This note proceeds as follows. Section \ref{section:theory} sets up the problem and presents the main theoretical results. Section \ref{section:implementation} shows how to feasibly implement the screened estimator and test for the reliability of stated compliance. Finally, Section \ref{section:conclusion} concludes with a brief discussion of future work.

%%%%%%%%%%%%%%
%%% THEORY %%%
%%%%%%%%%%%%%%
\section{Theory}\label{section:theory}
In this section, we describe the set up and present the main theoretical results. Proofs are in Appendix \ref{appendix:proofs}.

The main aim is to compare the standard \textit{unscreened IV estimator} to an alternative approach where experimenters screen out certain participants and then run the experiment on the remaining subset. We call this the \textit{screened IV estimator}. 

The theoretical results are proved under an idealized setting in which participant types are observed by the experimenter (i.e. complier, always-taker, never-taker). We derive an `optimal' screened estimator under this assumption. Since types are not observed in reality, we discuss how to implement this estimator in practice in Section \ref{section:implementation}.

\subsection{Setup}
Consider the standard linear IV model with heterogeneous treatment effects:
$$
Y_i = \alpha + \beta_i D_i + u_i
$$
$$
D_i = \phi + \pi Z_i + \eta_i
$$
\noindent
where $\mathbbm{E}[D_i u_i] \neq 0$ and we assume, for simplicity, that errors are homoscedastic. Let $S_i$ be a binary variable that equals one if $i$ is screened into the sample, and zero otherwise. We impose the standard assumptions under the LATE framework \citep{Imbens1994}:

\begin{assumption}[LATE Identifying Assumptions]\label{assumption:late}
Assume that
\begin{enumerate}  
    \item Exclusion restriction: $Y_i(d,0)=Y_i(d, 1) = Y_i(d)$ for $d=0,1$ 
    \item Independence: $\big(Y_i(D_i(1), 1), Y_i(D_i(0), 0), D_i(1), D_i(0),  S_i\big) \perp Z_i$  
    \item First Stage: $E[D_i(1) - D_i(0)] \neq 0$
    \item Monotonicity: $D_i(1)-D_i(0)\geq 0$ for all $i$ (or vice versa).
\end{enumerate}
\end{assumption}

Compared to the standard LATE assumptions, we include one additional assumption, namely, that the instrument is assumed to be independent of screening. This is immediately satisfied in the scenario we consider, where screening occurs prior to randomization.

The main aim is to compare median bias and power of the \textit{screened IV estimator}, $\hat{\beta}_r$, and the \textit{unscreened IV estimator}, $\hat{\beta}_1$, where the subscript denotes the fraction of the original sample that is retained. To make meaningful comparisons, we assume throughout that screening is neither trivial nor degenerate, in that at least one unit is screened out of the experiment and not all are screened out: $r \in (0,1)$. 

A fundamental concern for alternative IV estimators is whether or not they change the target parameter. For example, \citet{Spiess2021} and \citet{Hazard2025} propose alternative estimators that can provide gains in precision, but, in general, target a different LATE to standard IV estimation. This makes direct comparisons challenging. To avoid this issue, we establish a sufficient condition under which screening does not change the target parameter:

\begin{assumption}[Complier Retention]\label{assumption:complier_retention}
 $\mathbbm{P}[S_i=1 | CM_i = 1] = 1$
\end{assumption}

Under this sufficient condition, the target parameter is unchanged:

\begin{lemma}\label{lemma:invariance}
    Under Assumptions \ref{assumption:late} and \ref{assumption:complier_retention},
    $$
    \frac{\mathbbm{E}[Y_i|Z_i=1, S_i=1]-\mathbbm{E}[Y_i|Z_i=0, S_i=1]}{\mathbbm{E}[D_i|Z_i=1, S_i=1]-\mathbbm{E}[D_i|Z_i=0, S_i=1]}
    =
    \mathbbm{E}[Y_i(1)-Y_i(0)|  CM_i = 1]
    $$
\end{lemma}

In words, this result says that any screening procedure which retains all compliers will target the same LATE as the unscreened estimator (i.e. the average treatment effect on compliers). We maintain this assumption throughout the paper and propose a simple approach to test it in Section \ref{section:implementation}. 

\subsection{Optimal Screening Mechanism}
This subsection considers the optimal screening mechanism conditional on retaining all compliers (Assumption \ref{assumption:complier_retention}). The mechanism we propose is optimal in two senses: first, that it maximizes statistical power; and second, that it minimizes median bias. This differs from alternative IV estimators in the literature, which typically pose a bias-variance trade-off \citep{Spiess2021, Hazard2025}.

First, consider statistical power, for which we make the following assumption: 

\begin{assumption}[Homoscedasticity]\label{assumption:variance_stability}
$\mathbbm{E}[u_i^2] = \sigma_u^2$
\end{assumption}

The variance of the 2SLS estimator depends on the average residual variance in the estimation sample; under homoscedasticity, this average is unaffected by screening, ensuring that any reduction in the standard error comes solely from the strengthened first stage rather than from changes in error variance.\footnote{This can be weakened to allow for heteroscedasticity across units, provided that the average variance is equal across screened and unscreened samples. An even weaker requirement is that the screened sample have weakly lower average residual variance, $\sigma_{u,r}^{2} \le \sigma_{u,1}^{2}$, which also guarantees improved precision.} With this, we can state the main result:

\begin{proposition}\label{proposition:power}
    Under Assumptions \ref{assumption:late}--\ref{assumption:variance_stability}, $se(\hat{\beta}_{r}) =  \sqrt{r} \cdot se(\hat{\beta}_1)$ and $se(\hat{\beta}_{r})$ is minimized when all non-compliers are screened out.
\end{proposition}

This result shows that any (non-trivial) screening mechanism that retains all compliers lowers the standard error, which corresponds to a lower MDE and higher statistical power.\footnote{Statistical power is increasing in the ratio of the true treatment effect and the standard error. Since the true treatment effect is unchanged under Assumption \ref{assumption:complier_retention}, a lower standard error immediately implies higher statistical power.} Moreover, the power gain is inversely related to the fraction of the full sample that is retained, $r$. This implies that the power-maximizing screening mechanism minimizes $r$, which is achieved by screening out all never-takers and always-takers and retaining only compliers. 

The intuition for Proposition \ref{proposition:power} is that LATE identifies the effect for compliers only, so removing non-compliers reduces noise and thus increases power by `strengthening the instrument'. This general insight is well understood in the existing IV literature \citep{Spiess2021, Hull2025, Hazard2025}. The primary contribution of Proposition \ref{proposition:power} relative to existing work is therefore conceptual, in that it shows that this insight can be used to optimize \textit{experimental design}; this differs from the complementary but distinct aim of improving precision in \textit{existing} IV studies. 

To gauge the potential practical importance of this insight, consider \citet{tarozzi2015}, which studies the impact of microcredit in Ethiopia using an RCT with partial compliance. The first-stage shows that assignment to access to micro-credit increased borrowing rates by 25 percentage points; note that this also provides an estimate of the share of compliers in the population.\footnote{Formally, Assumption \ref{assumption:late} implies: $\pi = \mathbbm{E}[D_i | Z_i = 1] - \mathbbm{E}[D_i | Z_i = 0] = \mathbbm{E}[D_i(1) - D_i(0)] = \mathbbm{P}[CM_i=1]$.} 

Now suppose experimenters effectively screen out all non-compliers from the estimation sample. Then Lemma \ref{lemma:invariance} implies that the target parameter is unchanged. Moreover, under the assumptions of Proposition \ref{proposition:power}, the standard error for the screened estimator will be halved relative to the standard, unscreened estimator ($\sqrt{0.25} = 0.5$). Thus, confidence intervals and the MDE would also be halved, which represents an extremely large gain in precision for zero cost.

Next, we turn to bias, where we make the following assumptions:

\begin{assumption}[Independent Normal Errors]\label{assumption:normal_errors}
$u_i \overset{iid}{\sim} N(0, \sigma_u^2)$ and $\eta_i \overset{iid}{\sim}  N(0, \sigma_v^2)$ and $u_i \perp \eta_i$.
\end{assumption}

\begin{assumption}[Sign Screening on Estimated First Stage]\label{assumption:first_stage_screening}
    Experimenters screen for IV estimates where the sign of the estimated first-stage matches the sign of the population first-stage. 
\end{assumption}

Assumption \ref{assumption:normal_errors} imposes normality of errors for analytical convenience. Following \citet{Angrist2024}, Assumption \ref{assumption:first_stage_screening} assumes that experimenters screen the estimated first-stage to have the same sign as the population first-stage. In practice, this is a relatively mild assumption. 

To illustrate, consider again the example of a microcredit intervention. The population first-stage is generally thought to be positive ($\pi>0$), since those who are offered microcredit ($Z_i=1$) should have (weakly) higher take-up rates than those not offered ($Z_i=0$). Screening the estimated first-stage in this case means that if the estimated take-up rate were \textit{lower} in the group offered microcredit, then the experimenter would discard this quantitative results from the experiment. This is likely to hold in practice because such an outcome would likely invalidate the study since it suggests an extremely weak instrument. 

%\hl{ADD ASSUMPTION ABOUT equal fraction in treated and untreated in screened sample $N_0 = N (1-q)$. Reference in proof.}

With this, we can state the result:

\begin{proposition}\label{proposition:median_bias}
    Under Assumptions \ref{assumption:late},\ref{assumption:complier_retention}, \ref{assumption:normal_errors} and \ref{assumption:first_stage_screening}, the median of the magnitude of bias of the screened 2SLS estimator is strictly lower than that of the unscreened 2SLS estimator and minimised when all non-compliers are screened out.
\end{proposition}

This results state that any screening mechanism which retains all compliers lowers median bias -- and also that median bias will be lowest when all never-takers and always-takers are excluded. The intuition is similar to Proposition \ref{proposition:power}: since we only ever identify the impact on compliers, removing always-takers and never-takers effectively increases the strength of the instrument, which in turn lowers bias.

To summarize, this section makes a number of complementary claims. First, that a sufficient condition for maintaining the same target parameter in the screened and unscreened designs is to retain all compliers. Second, that conditional on satisfying this sufficient condition, screening out all always-takers and never-takers will minimize bias and maximize statistical power. Next, we turn to feasible implementation of this screening design.

\section{Feasible Implementation}\label{section:implementation}

In the preceding section, the theoretical results were derived assuming the experimenter can directly observe participants' types (e.g. complier, always-taker, never-taker) and screen accordingly. In reality, types are unobservable. This section discusses how to implement the optimal screening estimator in practice, and proposes an approach to partially test whether screening effectively identifies types. 

\subsection{Screening Questions} 
We recommend experimenters screen for compliers by simply asking them their intended compliance. Thus, for RCTs with two-sided noncompliance, we propose adding two questions to the questionnaire prior to randomisation:

\begin{enumerate}
    \item Would you take treatment $(D_i=1)$ if offered $(Z_i=1)$?
    \item Would you take treatment $(D_i=1)$ if not offered $(Z_i=0)$?
\end{enumerate}

Responses to these two questions identify participants' \textit{stated type}. For example, if a participant respond no to (i) and yes to (ii), then their \textit{stated type} is a complier. The effectiveness of screening depends on whether participants' \textit{stated type} match their \textit{true type}, which we discuss further in the next subsection. 

Before that, it is useful to introduce some notation. Let $\widetilde{CM}_i=1$ if $i$ is a \textit{stated complier} based on the two questions above, and zero otherwise; and let $CM_i=1$ if $i$ is a \textit{true complier} and zero otherwise. Additionally, we say that \textit{Type-I screening error} occurs when an individual is a stated complier ($\widetilde{CM}_i=1$) but not a true complier ($CM_i=0$). Alternatively, \textit{Type-II screening error} occurs when an individual is a stated non-compliers ($\widetilde{CM}_i=0$) but is in fact a true complier ($CM_i=1$). 

Identification requires that all compliers are retained; in other words, that there is no Type-II error. That, under screening for stated compliers ($S_i=1 \iff \widetilde{CM}_i=1$),  Assumption \ref{assumption:complier_retention} becomes $\mathbbm{P}[ \widetilde{CM}_i=1 | CM_i=1]=1$, which we discuss how to estimate in the next subsection. Note that Type-I error does not, by contrast, harm identification, although it does reduce the power gain because non-compliers included in the sample will increase the variance of the estimator. 

Finally, note that in RCTs with one-sided noncompliance only question (1) needs to be added to the questionnaire; this is because the answer to (2) is `no' by design for all participants.

\subsection{Pseudo-Screening and Testing Identification}
We recommend that experimenters implement what we call \textit{pseudo-screening}. This is where screening questions are asked to elicit stated types, but where the experiment nevertheless is conducted on the full sample. Pseudo-screening means that experimenters can calculate both the IV estimate and standard error for both the screened sample and the full sample, and allows experimenters to measure the power gains from the screened design.

More importantly, however, is that pseudo-screening provides experimenters with greater variation for testing the identification assumption, namely, that screening successfully retains all compliers, which ensure the same LATE as the unscreened estimator (Assumption \ref{assumption:complier_retention}). To see this, consider the case experimenters do not pseudo-screen but actually screened out stated non-compliers ($\widetilde{CM}_i=0$). Suppose further that a subset of stated non-compliers are in fact compliers ($CM_i=1$); this represents Type-II error, which violates the identification assumption. 

However, under true screening, this type of violation is undetectable because non-compliers are excluded from the experiment, and their behaviour is therefore never observed. By contrast, under pseudo-screening, a subset of stated non-compliers will be offered treatment, and their choices can be observed to test whether stated non-compliance matches non-compliance in reality.

%This can be tested, for example, by estimating the take-up rate for the `pseudo-screened-out' group. Further details on testing the identification assumption are discussed in the next subsection.

%\subsection{Testing the Identification Assumption}
%The theoretical results in Section \ref{section:theory} show that any screening mechanism that retains all compliers will identify the same LATE as the unscreened estimator (Assumption \ref{assumption:complier_retention}). In what follows, we show how we can test the validity of this assumption. 

To see how this can be formally tested, consider first the more general result in Proposition 1 in \citet{Hull2024}. This result can be used to show that for any characteristic $X_i$, the 2SLS coefficient from an IV regression of $X_i \times D_i$ on $D_i$ will estimate the average of that characteristic in the complier subpopulation i.e. $\mathbbm{E}[X_i | CM_i=1]$. 

Returning to our setting, let $\widetilde{CM}_i$ be a binary variable that equals one if $i$ reports the intention to comply, and zero otherwise. From the general result above, it follows that an IV regression of $\widetilde{CM}_i \times D_i$ on $D_i$ will estimate the probability that stated compliance equals one conditional on actually being a complier i.e. $\mathbbm{P}[ \widetilde{CM}_i=1 | CM_i=1]$. This corresponds exactly the probability in Assumption \ref{assumption:complier_retention} when the screening mechanism is to retain all stated compliers. 

Thus, we propose a one-sided test of the null hypothesis that $\mathbbm{P}[ \widetilde{CM}_i=1 | CM_i=1]=1$, where the alternative hypothesis is $\mathbbm{P}[ \widetilde{CM}_i=1 | CM_i=1]<1$.\footnote{An alternative test of the identification assumption would be to estimate whether treatment take-up is zero for the `pseudo-screened-out' group (i.e. stated non-compliers). We opt for the test in the main text because it utilizes the full sample and is therefore more powerful.} Rejecting the null hypothesis suggests that stated compliance is not reliable. This might justify using of the standard, unscreened estimator, which is, of course, calculable under pseudo-screening. On the other hand, failing to reject the null hypothesis would favor using the screened estimator, as this is guaranteed to have better statistical power (Proposition \ref{proposition:power}) and lower bias (Proposition \ref{proposition:median_bias}).

\subsection{Optimal Screening Estimator}
Validating that $\mathbbm{P}[ \widetilde{CM}_i=1 | CM_i=1]=1$ tests for Type-II screening error; and hence it provides a direct test on whether the LATE is preserved and that the screen estimator has better bias and power properties than the unscreened estimator. However, it does not imply that there is no Type-I screening error, and hence whether the optimal screening design has been achieved. That is, some participants may state being compliers ($\widetilde{CM}_i=1$) but may not in fact be compliers $CM_i=0$. This outcome might arise, for example, due to Hawthorne effects.
More formally, it is possible that $\mathbbm{P}[ \widetilde{CM}_i=0 | CM_i =0] \neq 1$, even if $\mathbbm{P}[ \widetilde{CM}_i=1 | CM_i =1]=1$. 

Thus, to test whether the optimal screening design was obtained, the experimenter can additionally test the null hypothesis that $\mathbbm{P}[ \widetilde{CM}_i=0 | CM_i=0] = 1$. To calculate this quantity, note that the law of total probability implies the following:
\[
\mathbb{P}\!\left[\widetilde{CM}_i = 0 \mid CM_i = 0\right]
=
\frac{
\mathbb{P}\!\left[\widetilde{CM}_i = 0\right]
-
\mathbb{P}(CM_i = 1)\,
\mathbb{P}\!\left[\widetilde{CM}_i = 0 \mid CM_i = 1\right]
}{
1 - \mathbb{P}[CM_i = 1]
}
\]

\noindent
where all quantities on the right-hand side are directly estimatable. 

Hence, if both nulls are not rejected, then this provides evidence that stated responses are reliable and that the optimal screening estimator is achieved.

\section{Conclusion}\label{section:conclusion}
This note examines optimal experimental design under partial compliance when experimenters can screen participants prior to randomization. It shows that screening for compliers maintains the same LATE as no screening but maximizes statistical power and minimizes bias. Back-of-the-envelope calculations based on actual RCTs suggest that power gains could be quite large e.g. halving confidence intervals.

However, the feasibility of the screened estimator depends on the reliability of eliciting compliance behaviour. Thus, we are pursuing ongoing work to test the feasibility and potential advantages of the optimal screening design over a standard compliance design.

\newpage
\bibliographystyle{aer}
{\small
\bibliography{References}

@article{Imbens1994,
  author    = {Imbens, Guido W. and Angrist, Joshua D.},
  title     = {Identification and Estimation of Local Average Treatment Effects},
  journal   = {Econometrica},
  volume    = {62},
  number    = {2},
  pages     = {467--475},
  year      = {1994},
  publisher = {Econometric Society},
  doi       = {10.2307/2951620}
}

@article{Hazard2025,
  author    = {Hazard, Yagan and L\"owe, Simon},
  title     = {Improving Late Estimation in Experiments with Imperfect Compliance},
  journal   = {Working Paper},
  volume    = {},
  number    = {},
  pages     = {},
  year      = {2025},
  publisher = {},
  doi       = {}
}

@article{Spiess2021,
  author    = {Spiess, Jann and Coussens, Stephen},
  title     = {Improving Inference from Simple Instruments through Compliance Estimation},
  journal   = {Working Paper},
  volume    = {},
  number    = {},
  pages     = {},
  year      = {2021},
  publisher = {},
  doi       = {}
}

@article{tarozzi2015,
  title        = {The Impacts of Microcredit: Evidence from Ethiopia},
  author       = {Tarozzi, Alessandro and Desai, Jaikishan and Johnson, Kristin},
  journal      = {American Economic Journal: Applied Economics},
  volume       = {7},
  number       = {1},
  pages        = {54--89},
  year         = {2015},
  doi          = {10.1257/app.20130475},
  url          = {https://www.aeaweb.org/articles?id=10.1257/app.20130475}
}

@article{Angrist2024,
  title        = {One Instrument to Rule Them All: The Bias and Coverage of Just-ID IV},
  author       = {Angrist, Joshua D. and Koles{\'a}r, Michal},
  journal      = {Journal of Econometrics},
  year         = {2024},
  month        = mar,
  volume       = {240},
  number       = {2}
}

@article{Hull2025,
  title        = {Optimal Formula Instruments},
  author       = {Hull, Peter and Borusyak, Kirill},
  journal      = {Working Paper},
  year         = {2025},
  volume       = {},
  number       = {}
}

@article{Hull2024,
  author    = {Borusyak, Kirill and Hull, Peter},
  title     = {Negative Weights Are No Concern in Design-Based Specifications},
  journal   = {AEA Papers and Proceedings},
  volume    = {114},
  pages     = {597--600},
  year      = {2024},
  doi       = {10.1257/pandp.20241062},
  url       = {https://www.aeaweb.org/articles?id=10.1257/pandp.20241062},
}
}

%%%%%%%%%%%%%%%%
%%% APPENDIX %%%
%%%%%%%%%%%%%%%%
\newpage
\appendix

\section{Proofs}\label{appendix:proofs}
\setcounter{table}{0}
\setcounter{figure}{0}
\noindent \textbf{Proof of Lemma \ref{lemma:invariance}}:
    Observe that 
    \begin{align*}
    \frac{\mathbbm{E}[Y_i|Z_i=1, S_i=1]-\mathbbm{E}[Y_i|Z_i=0, S_i=1]}{\mathbbm{E}[D_i|Z_i=1, S_i=1]-\mathbbm{E}[D_i|Z_i=0, S_i=1]}
    = & 
    \mathbbm{E}[Y_i(1)-Y_i(0)|  CM_i = 1, S_i=1] \\
    = & \mathbbm{E}[Y_i(1)-Y_i(0)|  CM_i = 1]
    \end{align*}
    The first equality follows from the arguments in the standard LATE theorem with Assumption \ref{assumption:late} conditional on $S_i=1$ \citep{Imbens1994}. The second equality follows immediately from Assumption \ref{assumption:complier_retention} \qed

\bigskip
\noindent \textbf{Proof of Proposition \ref{proposition:power}}: The ratio of the standard error under the screened design relative to the standard error under the standard design is, under homoscedasticity, 
    \begin{align}\label{equation:se_ratio}
    \frac{se(\hat{\beta}_r)}{se(\hat{\beta_1})} = 
    \sqrt{
    \frac{
    \frac{\sigma_{u,r}^2}{N_S \cdot \pi_S^2 \cdot \widehat{Var}(Z)
    }}{
    \frac{\sigma_{u,1}^2}{N \cdot \pi^2 \cdot \widehat{Var}(Z)}
    }
    }
    =
    \sqrt{\frac{N}{N_S}} \cdot\frac{\pi}{\pi_S}
    \end{align}
    \noindent
    where the final inequality uses $\sigma_{u,S}^2 = \sigma_{u}^2$ (Assumption \ref{assumption:variance_stability}). 
    
    Next, note that Assumption \ref{assumption:late} implies $\pi = \mathbbm{E}[D_i | Z_i = 1] - \mathbbm{E}[D_i | Z_i = 0] = \mathbbm{E}[D_i(1) - D_i(0)] = \mathbbm{P}[CM_i=1]$. By a similar argument, we have
    \begin{align*}
    \pi_S = \mathbbm{P}[CM_i=1| S_i=1] 
    = &
    \frac{\mathbbm{P}[S_i=1 | CM_i=1]\mathbbm{P}[CM_i=1]}
    {\mathbbm{P}[S_i=1]} \\
    = & \frac{\mathbbm{P}[CM_i=1]}{r}
    \end{align*}
    \noindent
    where the first equality uses Bayes Rule. It follows that $\pi/\pi^S = r$. Finally, since $N_r = r N$ by definition, it we have that $\sqrt{\frac{N}{N_S}} = \frac{1}{\sqrt{r}}$. Substituting these expressions into equation \eqref{equation:se_ratio} gives the desired result.

\bigskip
\noindent \textbf{Proof of Proposition \ref{proposition:median_bias}}:
    Define for convenience $q \equiv \frac{1}{N} \sum_i Z_i$. We begin by deriving an expression for bias of the screened 2SLS estimator. 
    $$
        \hat{\beta}_r - \beta 
        = \frac{\sum_{\{i \mid S_i=1\}} \tilde{Z}_i u_i}{\sum_{\{i \mid S_i=1\}} \tilde{Z}_i \tilde{D}_i} 
    $$

    Note that this expression also captures the unscreened estimator, $\hat{\beta}_1$, when $r=1$. Expanding the denominator gives
    \begin{align*}
    \sum_{\{i \mid S_i=1\}} \tilde{Z}_i \tilde{D}_i = & \sum_{\{i \mid S_i=1\}} \tilde{Z}_i {D}_i - \bar{D} \sum_{\{i \mid S_i=1\}} \tilde{Z}_i \\
    = & \sum_{\{i \mid S_i=1\}} \tilde{Z}_i {D}_i = \sum_{\{i \mid S_i=1\}} \tilde{Z}_i (\phi + \pi_S Z_i + \eta_i) \\
    = & \pi_S \sum_{\{i \mid S_i=1\}} \tilde{Z}_i Z_i + \sum_{\{i \mid S_i=1\}} \tilde{Z}_i \eta_i \\
    = & \pi_S N_S q(1-q) + \sum_{\{i \mid S_i=1\}}  \tilde{Z}_i\eta_i \\
    = & \pi N q(1-q) + \sum_{\{i \mid S_i=1\}}  \tilde{Z}_i\eta_i 
    \end{align*}
    The second and fourth equalities use the fact that $\sum_{\{i \mid S_i=1\}} \tilde{Z}_i = \sum_{\{i \mid S_i=1\}} (Z_i-q) = \sum_{\{i \mid S_i=1\}} Z_i - rNq = 0$. The final line uses the following simplification.
    \begin{align*}
    \pi_S = & \mathbbm{E}[D_i | Z_i=1, S_i=1] - \mathbbm{E}[D_i | Z_i=0, S_i=1] \\
     = & \mathbbm{E}[D_i(1) - D_i(0) | S_i=1] \\
      = & \mathbbm{E}[CM_i = 1 | S_i=1] \\
      = & \mathbbm{P}[CM_i = 1 | S_i=1]] \\
      = & \frac{\mathbbm{P}[S_i=1 | CM_i = 1 ] \mathbbm{P}[CM_i=1]}{\mathbbm{P}[S_i=1 ]} =  \frac{\pi}{r}
    \end{align*}
    
    \noindent
    where the second equality uses independence (Assumption \ref{assumption:late}); the third equality monotonicity (Assumption \ref{assumption:late}); the fifth equality Bayes Rule; and the final equality follows due to full complier retention (Assumption \ref{assumption:complier_retention}), the fact that $r \equiv \mathbbm{P}[S_i=1]$, and finally that $\pi =\mathbbm{P}[CM_i=1]$. Since $N_S = r N$, it follows that $\pi_S N_S = \pi N$.

    Thus, we have
    \begin{align*}
    \hat{\beta}_r - \beta = \frac{
            M_{u, r}
            }{
            \mu + M_{\eta, r}
            } \label{equation:bias_unscreened}
    \end{align*}
    \noindent
    where $\mu \equiv \pi Nq(1-q)$, $M_{\eta, r} \equiv \sum_{\{i \mid S_i=1\}} \tilde{Z}_i \eta_i$, and $M_{u, r} \equiv \sum_{\{i \mid S_i=1\}} \tilde{Z}_i u_i$.

    Next, observe that $M_{u, r}$ and $M_{\eta, r}$ are linear combinations of independent Gaussian draws, from which it follows that
    \begin{align*}
    \text{Var}(M_{u, r}) = & \text{Var}\left(\sum_{\{i \mid S_i=1\}}  \tilde{Z}_iu_i \right) 
    = \sum_{\{i \mid S_i=1\}} \text{Var}\left( \tilde{Z}_iu_i \right)  \\
      = & \sigma_u^2\sum_{\{i \mid S_i=1\}} \tilde{Z}_i^2 = r\cdot \left( \sigma_u^2 N q(1-q)\right)\\
      = & r\cdot \text{Var}(M_{u, 1})
    \end{align*}

    In words, the variance of $M_{u, r}$ is a scaled version of the corresponding unscreened random variable, $M_{u, 1}$. By the same argument $\text{Var}(M_{\eta, r}) = r\cdot \text{Var}(M_{\eta, 1})$. It follows that
    \begin{align}
    \hat{\beta}_r - \beta = \frac{
            M_{u, r}
            }{
            \mu + M_{\eta, r}
            } \overset{d}{=} 
            \frac{
            \sqrt{r} M_{u, 1}
            }{
            \mu + \sqrt{r} M_{\eta, 1}
            }
    \end{align}

    Let $F_{Bias| \eta}(\cdot)$ the cdf of bias conditional on a fixed value of $M_{\eta,r}$; and let  $\Phi(\cdot)$ denote the normal cdf. Then for any $t>0$
    \begin{align*}
        F_{Bias | r, \eta}(-t) = & \mathbbm{P}\left(
          \frac{\sqrt{r} M_{u, 1} }{\mu + \sqrt{r} \cdot m}  
          < -t   \; \Bigg| \; M_{\eta, 1} = m
        \right) \\
         = & \mathbbm{P}\left(
           M_{u, 1} < -\frac{t\mu}{\sqrt{r}} - t \cdot m  \; \Bigg| \; M_{\eta, 1} = m
        \right) \\
         = & \Phi\left(
          -\frac{t\mu}{\sqrt{r}} - t \cdot m
        \right)
    \end{align*}
    \noindent
    where the second equality uses the fact that the denominator is positive, which occurs precisely when the estimated first-stage has the same sign as the population first stage (Assumption \ref{assumption:first_stage_screening}). The final expression is clearly increasing in $r$.

   By similar reasoning, we have that $1- F_{Bias | r, \eta}(-t) = 1- \Phi\left(\frac{t\mu}{\sqrt{r}} + t \cdot m \right)$, which is also increasing in $r$. 
   
   Thus, the probability mass of the tails of the \textit{conditional} bias distribution are increasing in $r$. Next, note that a straightforward application of the dominated convergence theorem implies that the probability mass of the tails of the \textit{unconditional} bias distribution are also increasing in $r$. 
   
   Finally, let $F_{|Bias|, r}(\cdot)$ denote the cdf of the absolute value of bias. From the preceding arguments, it follows that for any $0 < r_1 < r_2 <1$, and $t>0$, that
   \begin{align*}
        F_{|Bias|, r_2}( t) \leq F_{|Bias|, r_1}( t)
   \end{align*}
   In other words, $F_{|Bias|, r_2}(\cdot)$ first-order stochastically dominates $F_{|Bias|, r_1}(\cdot)$, which immediately implies that the median (and in fact all other quantile) under $F_{|Bias|, r_2}( \cdot)$ is larger than under $F_{|Bias|, r_1}(\cdot )$, which is what we wanted to show. 

   From this, it is immediately clear that the bias minimizing choice for $r$ is to screen out all never-takers and always-takers, such that $r = \mathbbm{P}(CM_i=1)$. 
\qed

\begin{comment}
\section{Power Gain When Maintaining Sample Size}
\setcounter{table}{0}
\setcounter{figure}{0}
Suppose the experimenter has a fixed budget that allows them to conduct the experiment on $N$ individuals. The optimal design proposed in the main text would collect data on $r N < N$ individuals, which would leave unutilized funds in the budget. This appendix consider gains to power and decreases in median bias in the case where the experimenter fully utilizes their budget by collecting additional data such that the final, screened dataset is size $N$. 
\end{comment}

\end{document}